# Joint Load and Capacity Scheduling for Flexible Radio Resource Management of High-Throughput Satellites

Jia Zhuoya, Xiong Wei *, Hao Hongxing, Liu Zheng, Han Chi

**Abstract:** This work first explores using flexible beam-user mapping to optimize the beam service range and beam position, in order to adapt the non-uniform traffic demand to offer in high-throughput satellite (HTS) systems. Second, on this basis, the joint flexible bandwidth allocation is adopted to adapt the offer to demand at the same time. This strategy allows both beam capacity and load to be adjusted to cope with the traffic demand. The new information generated during the load transfer process of flexible beam-user mapping can guide the direction of beam optimization. Then, the proposed strategies are tested against joint power-bandwidth allocation and joint optimization of bandwidth and beam-user mapping under different traffic profiles. Numerical results are obtained for various non-uniform traffic distributions to evaluate the performance of the solutions. Results show that flexible joint load and capacity scheduling are superior to other strategies in terms of demand satisfaction with acceptable complexity. Our source code along with results are available at crystal-zwz/HTS_RRM_Joint-Load-and-Capacity-Scheduling (github.com).

## 1 Introduction

Next–generation satellite communication (SatCom) networks can facilitate escalating data rates for seamless incorporation into expansive satellite-terrestrial networks. The flexible Ultra High Throughput Satellite (UHTS) systems can offer Terabit per second in–orbit capacity when and where needed[1]. Currently, more sophisticated HTS equipped with adaptive resource assignment capabilities are increasingly gaining popularity. For instance, the recently launched ViaSat-3 constellation is designed to bring bandwidth when and where needed most; SES-17 and O3b mPOWER use Adaptive Resource Control (ARC) software to exploit the flexibility of digital payloads, concentrating power in specific beams as needed. These advanced configurations are designed to unlock new applications in robotics, wireless security, and the Internet of Things (IoTs) including vast heterogeneous devices, with cost-effective ubiquitous connectivity and provide a competitive alternative to the future generation network. Novel services will generate time-varying and un-uniformly distributed traffic demand, leading to even larger fluctuations of this heterogeneous demand[2]. In response to the above challenges, the ability to swiftly and flexibly assign radio resources in alignment with the dynamic demand pattern has emerged as an essential requirement[3].

While full digital payload combined with software-defined Radio Resource Management (RRM) is developed enough, the algorithms necessary to fully leverage such technology are at an early stage of maturity. Over recent years, extensive research efforts have focused on enhancing resource utilization efficiency by exploring all available degrees of flexibility, including optimization in time, frequency,

● Jia Zhuoya, Xiong Wei *, Hao Hongxing, Liu Zheng, Han Chi are with the National Key Laboratory of Space Target Awareness, Space Engineering University, Beijing 101416, China. E-mail: zorajade@hgd.edu.cn, wei_xiong@hgd.edu.cn, hongxinghao87@hgd.edu.cn, liuzheng@hgd.edu.cn, hanchi@hgd.edu.cn.
* To whom correspondence should be addressed.

spatial domains and power allocation, cf.[3-8]. Yet, during the design stage, there is often less emphasis on the significant interdependencies between subsystems. For example, the utility of the joint optimization of reconfigurable active antenna and flexible repeater rarely considered. Should critical design issues not be addressed collaboratively with all involved subsystems, no matter how well each subsystem might be optimized individually, the holistic system solution would be defective in terms of costs and performance metrics.

In recent years, there has emerged a focus on the joint design of flexible antenna. Argyrios Kyrgiazos et al.[9] first introduced a joint optimization of bandwidth allocation and beam sizes to match the uneven traffic demands. Jean-Thomas Camino et al.[10] emphasized that the optimization of the beam layouts occurs early in system design, thus any shortcomings from this stage cannot be remedied for later. But [9, 10] just defined a limited number of beam sizes to simplify the model. Piero Angeletti et al.[11] presented a methodology generating Voronoi diagram iteratively until load balance is achieved, mainly driven by load gradients. However, the existence of the solution depends entirely on the number of beams and the demand profile, and the iteration step size needs to be chosen carefully, otherwise, the algorithm would diverge. In conclusion, none of these methods fully take advantage of advanced digital payloads, as increased payload flexibility leads to a significant increase in algorithmic complexity. If we want to utilize advanced payload with less complexity, the flexibility when scheduling resources in the satellite downlink can be achieved through a beam-free approach, in which the beam-user mapping can be adjusted on demand[12].

This work aims to fully utilize the advanced HTS's flexibilities to perform the RRM. In short, the main contributions are summarized below:

(1) A novel user-centric resource management framework is presented, which balances beam load through a two-way effort on both the supply side and the demand side--the active antenna adapts to mitigate the unevenness in demand, thereby alleviating the challenge faced by the flexible repeater to meet the requirement.

(2) By introducing flexible beam-user mapping，we transform the original non-convex problem into a simple stepwise convex optimization. Starting from an initial uniform beam pattern, the footprints are iteratively displaced and resized, joint with flexible bandwidth, in a way that ultimately equalizes the traffic demand. The existence of the solution is independent of system parameters and traffic profile, without any coefficients needing to be adjusted manually.

(3) The performance of the proposed strategies is analyzed by comparing the performance with typical strategies presented in the last five years, including a joint power-bandwidth allocation and a joint optimization of bandwidth and beam-user mapping. The different traffic demand scenarios are modeled by the Dirichlet distribution and a more realistic traffic profile based on population density.

The sections of the paper are organized as follows. First, the flexible RRM system model and resource scheduling problem are introduced in Section II. Next, the optimization strategy embodying two-way effort is proposed in Section III. Section IV presents and compares the performance evaluation of the relevant strategies. Finally, section V concludes the paper.

## 2  Flexible RRM System Model

We consider the forward link of a satellite communication system (gateway-satellite-user), which consists of a geostationary (GEO) satellite that illuminates a $K = K_1 \times K_2$ array as part of its coverage. It is assumed that flexible antenna radiation is equipped to provide adaptable beam direction, yet the shape and size of each beam is fixed. We define the number of users served at a given moment in the given array is $N$, and the set of users served by the beam $k$ is denoted by $\mathcal{S}(k)$.

The flexible power allocation was reported to provide limited gain, compared to the bandwidth flexibility, due to the logarithmic relationship between power and capacity determined by Shannon's formula[13]. Therefore, considering a flexible architecture, the satellite payload can optimize the bandwidth allocated to different users and the beam direction. Moreover, the gateway can freely manage the beam-user mapping so that users are not constrained by their dominant serving beam. This means that the flexible mapping allows the user to pair with an adjacent non-dominant beam when the signal radiation intensity of the adjacent beams is sufficient.

Considering a four-color reuse pattern, the total system bandwidth is $W^{total}$. It is assumed that $N_{TWTA}$ traveling-wave tubes amplifiers (TWTAs) are available and each of them amplifies the whole bandwidth $W^{total}$. Subsets of beams connected to different TWTAs are disjoint, with neglected inter-beam interference. In each TWTA, the bandwidth is flexibly allocated to the subset of the beam connected to it, and the same part of the spectrum cannot be allocated to multiple beams connected to the same one-line wave tube, and the bandwidth is allocated to the carrier within the beam without overlapping. The portions of traffic addressed to the users served by a given carrier are multiplexed in time by time division multiplexing.

The flexible bandwidth distribution depends on digital channel and subband exchange technology. With the flexible repeater, the bandwidth can be divided disproportionally among the beams sharing the same TWTA. Assume that the pair $j$ of two consecutive beams, with indexes $A(j) = \{2j-1, 2j\}$, $j = 1,\ldots, K/2$ is served by the same amplifier with no bandwidth overlapping. The bandwidth can be adjusted in increments of $\tilde{W}$ MHz, i.e., the bandwidth allocated to a beam must be a multiple of $\tilde{W}$ MHz. This increment of bandwidth can be referred to as *chunk*. The available bandwidth $W^{total}$ is divided into $2M$ carriers with a fixed carrier bandwidth $\tilde{W} = W^{total}/2M$. Therefore, for $M_k$ carriers assigned to the beam $k$, the corresponding bandwidth is $M_k \tilde{W}$ MHz. Given the beam $k$ bandwidth, the number of carriers in any beam is

$$M_k = \left\lfloor \frac{W_k}{\tilde{W}} \right\rfloor \forall k.$$

The antenna pattern is represented by Bessel model. The channel gain from the *b*-th beam to the *n*-th user can be represented by

$$g_k(n) = G_{max} \left( \frac{J_1(u)}{2u} + 36 \frac{J_3(u)}{u^3} \right)^2 \quad (1)$$

with $u \approx 2.07123 \cdot d_b(n)/R$, and $d_b(n)$ the distance between peak gain axis of the *k*-th beam and the *n*-th user position. The end-to-end channel response from beam $k$ to the *n*th user is indicated by $h_k(n)$, with the corresponding magnitude denoted as

$$|h_k(n)| = \frac{\sqrt{G_R g_k(n)}}{\sqrt{k_B T \tilde{W} 4\pi D/\lambda}} \quad (2)$$

where $G_R$ represents the receive antenna gain, $\lambda$ denotes the carrier wavelength, and $k_B$, $T$ refer to the Boltzmann constant and the clear sky noise temperature, respectively. Regarding the layout of the beams, each one is spaced $2R$ apart from its neighbors, with the beam radius $R$ according to the Bessel model.

The primary goal of the system is to align the available offered rates $R^{off}(n)$ to the user demanded rates $R^{req}(n)$. Variations of this approach are widely found in the existing works[13-16]. The quadratic unmet demand is used since it offers a good balance between demand fulfillment and user fairness. The quadratic unmet objective function is expressed as

$$\mathcal{U} = \sum_{n=1}^{N} (R^{req}(n) - R^{off}(n))^2 \quad (3)$$

The offered rate to the user $n$ served by beam $k$ is expressed as

$$R^{off}(n) = \sum_{c=1}^{M_k} \omega_{c,n} \tilde{W} \log_2 \left( 1 + \frac{\tilde{p}_k |h_k(n)|^2}{N_0 \tilde{W}} \right), n \in \mathcal{S}(k) \quad (4)$$

where $\omega_{c,n}$ denotes the time slot that user $n$ uses the carrier $c$ in the beam $k$. $N_0$ is the noise power spectral density. Moreover, $u_{c,n}$ denotes whether or not the user $n$ is active in the carrier $c$, formulated as

$$\sum_{c=1}^{M_k} u_{c,n} \leq \zeta_{max}, \forall n \in \mathcal{S}(k) \quad (5)$$

where $\zeta_{max}$ is the maximum number of carriers, and $\zeta_{max} = 1$ for the single-carrier operation of the receive terminals. Moreover, the sum of $\omega_{c,n}$ in the beam $k$ serving all its users is less than 1, formulated as

$$\sum_{n \in \mathcal{S}(k)} \omega_{c,n} \leq 1 \forall k, c = 1,\ldots, M_k \quad (6)$$

We define a set of indicator variables $\{x_{k,n}\}$ to describe each user's group. If a user $n$ belongs to group $k$, $x_{k,n} = 1$; otherwise $x_{k,n} = 0$. Each user must be served once and only once for higher efficiency. Then we have

$$\sum_{k=1}^{K} x_{k,n} = 1, \forall n \quad (7)$$

All users in the same group should be located within the main lobe of a beam to be served at the same time. The shape of the beam service area is assumed to be circular. Thus, the constraint related to the coordinates of users can be expressed as

$$\sum_{k=1}^{K} x_{k,n} \|l_n - c_k\| \leq \sum_{k=1}^{K} x_{k,n} r_k, \forall n \quad (8)$$

where $c_k$ represents the coordinate of the beam center, $\|l_n - c_k\|$ is the distance between user $n$ and the center of group $k$, and $r_k$ is the service radius of each beam which is adjustable. Under the system parameters set in this study, it was reported that the current DVB-S2X standard can tolerate a 6.3dB reduction in the signal-to-noise (SNR) ratio for users served by a non-dominant beam, which requires the user to be less than $r_f$ away from the beam center, $r_k \leq r_f$ [16].

Based on the analysis above, we would like to solve the following problem:

$$\min_{W, \Upsilon, X, \Omega, U} \mathcal{U}$$

$$\text{s.t. } R^{off}(n) = \sum_{k=1}^{K} \upsilon_{n,k} \log_2 \left( 1 + \frac{\tilde{p}_{n,k} |h_k(n)|^2}{N_0 \upsilon_{n,k}} \right)$$

$$C1: W_k + W_{k+1} \leq W^{total}, \forall k$$

$$0 \leq W_k \leq W^{total}$$

$$C2: \sum_{k=1}^{K} \upsilon_{n,k} \leq \tilde{W}, \forall n$$

$$W_k = \sum_{n=1}^{N} \upsilon_{n,k}$$

$$v_{n,k} \geq 0$$

$$C3: \sum_{k=1}^{K} x_{k,n} \|l_n - c_k\| \leq \sum_{k=1}^{K} x_{k,n} r_k, \forall n, k$$

$$\sum_{k=1}^{K} x_{k,n} = 1, \forall n$$

$$x_{k,n} \in \{0,1\} \forall k, n$$

$$r_k \leq r_f, \forall k \qquad (9)$$

$$C4: \sum_{c=1}^{M_b} u_{c,n} \leq 1, \forall n \in \mathcal{S}(k)$$

$$\sum_{n \in \mathcal{S}(k)} \omega_{c,n} \leq 1, \forall k, c = 1,...,M_k$$

$$0 \leq \omega_{c,n} \leq u_{c,n} \leq 1$$

$$u_{c,n} \in \{0,1\} \forall k, c, n \in \mathcal{S}(k)$$

where $\boldsymbol{W} = [W_1,...,W_K] \in \mathbb{R}^K$ gathers the bandwidth assignment per beam. the power $p_{n,k}$ allocated to the $n$th user in beam $k$ is proportional to that of the bandwidth $v_{n,k}$, which is gathered in $\boldsymbol{\Upsilon} = [v_{1,1},...,v_{N,1},v_{1,2},...,v_{N,K}] \in \mathbb{R}^{N \cdot K}$. $\boldsymbol{\Omega}_k = [\omega_{1,1},...,\omega_{M_k,1},\omega_{1,2},...,\omega_{M_k,|\mathcal{S}(k)|}] \in \mathbb{R}^{M_k \cdot |\mathcal{S}(k)|}$ and $\boldsymbol{U}_k = [u_{1,1},...,u_{M_k,1},u_{1,2},...,u_{M_k,|\mathcal{S}(k)|}] \in \mathbb{R}^{M_k \cdot |\mathcal{S}(k)|}$ gather the carrier assignment for beam $k$, respectively grouped in $\boldsymbol{\Omega} = [\boldsymbol{\Omega}_1,...,\boldsymbol{\Omega}_K]$ and $\boldsymbol{U} = [\boldsymbol{U}_1,...,\boldsymbol{U}_K]$; $\boldsymbol{X} = [x_{1,1},...,x_{K,1},x_{1,2},...,x_{K,N}] \in \mathbb{R}^{K \cdot N}$ gathers the beam-user mapping. Constraint $C1$ ensures that the bandwidth used by two adjacent beams does not overlap. Constraint $C2$ limits the bandwidth allocated to a given user less than that of a carrier. $C3$ defines the beam-user mapping and the range of users served by the beam. Finally, $C4$ constrains the terminals to operate in single carrier mode.

If the problem constraint contains only $C1$ and $C2$, but not $C3$ and $C4$, then it is a convex Quadratic Programming (QP) problem, the outcome will be the beam-user mapping. The discrete variables $x_{k,n}$ and norm calculations in $C3$ contribute to a non-convex mixed integer programming problem. When only one beam is considered, the problem can be converted to the Smallest Enclosing Circle (SEC) problem[17, 18], whose object is to find the smallest circle that can cover a set of points in the plane. Because the SEC problem is NP-hard, this problem is also NP-hard. In the existing literature, greedy algorithm and random algorithm[19] are generally used to solve this kind of user grouping and beam service range optimization problem. However, with the number of beams increasing, the algorithm complexity will increase by cube.

We follow a two-step approach in literature to handle $C4$ in the next section. First optimize the resources across beams, including $C1$, $C2$ and $C3$, i.e. bandwidth, beam-user mapping, beam service range and position; then intra beam, just handling $C4$, i.e. the carrier power and number of carriers per beam.

## 3 Optimization Strategy Embodying Two-Way Effort

It can be assumed that a system is optimal if and only if it fails to meet any further demand after using up all available capacity. Conversely, a system is suboptimal if it leaves demand unmet while still having spare capacity. In this case, the system is not operating at its full potential, and resources are wasted due to imbalances in the distribution of loads. To optimize the system, we can pursue two main strategies: transferring resources from underloaded beams to overloaded ones to match the pattern of demand more closely; Spreading the load more evenly among the beams by redistributing demand away from overloaded ones to those with spare capacity. These two strategies can supplement each other, thereby achieving a more balanced and efficient system operation.

### 3.1 Joint Load and Capacity Scheduling

Our goal is to pose a tractable implementation for joint load and capacity scheduling in the resource management across beams. Let us revisit this flexible architecture:

In flexible beam-user mapping satellite systems, user terminals give back the strength of their two downlink strongest channels to the gateway, then the users can be assigned to the link with better quantity[16, 20, 21]. Therefore, a given user will be served not only by its dominant beam, but also the strongest neighbor beam. As a result, the load at the edge of overloaded beams can be transferred to their adjacent underloaded beams, making the demand tend to adapt to the offer than before. Notably, this improvement is operated cooperatively. Since it is a system-level management considering global metrics, all the beams involved in the resource pulling supply the traffic demand, even some of them far away from the hot spot.

Significantly, in the process of transferring load through flexible beam-user mapping, the orientation of user grouping optimization is reflected compared to the original state, i.e., within the allowable SNR tolerance, the underloaded beams tend to serve a larger geographical area with more users, while the overloaded beams tend to serve a smaller geographical area with fewer users. This preliminary trend is the new information generated when operating flexible mapping. It is easy to think that if utilizing this information to make further user grouping, then adjusting the beam service range according to the location of each group of users, this guided optimization strategy will obviously be more efficient than the random search for the beam center and user group.

Therefore, we should change the perspective of problem-solving and use this new information to transform constraint *C*3. After the beam-user mapping has been adjusted, the center of the smallest circle covering all users served by one beam can be the new beam center, and the radius of this smallest circle can be the new service radius. This approach has better fairness than using the mean of all user coordinates in the service area as the center of the circle because the wider beam will cause the edge user to receive less power. After the beam center and service range adjustments have been completed, it is natural to implement the next beam-user mapping optimization. Iterate this process until the load balancing degree is no longer decreasing. Define load balancing degree as a measure of whether to stop iteration:

$$\kappa = \sum_{k=1}^{K} \frac{\left| R_k^{req} - R^{design} \right|}{R^{design}} \quad (10)$$

where $R^{design}$ is the design capacity of each beam.

Based on the above analysis, we propose a simple iteration as a practical implementation for the complex optimization problem caused by *C*3. The iterative algorithm is summarized in Algorithm 1. For a given number of beams, solve relaxed P1 by iterations. The starting point of the algorithm is the resource allocation used in the conventional payload. In each iteration, first, we calculate the current load balancing degree by equation and the current SNR matrix between each user and each beam center. Then we solve the QP problem to get the $R^{off}$ matrix, which contains the two largest rates that the user can get. By choosing the beam index corresponding to the maximum offer rate, then we obtain a new beam-user mapping. The bandwidth allocated to each beam is then discretized to get the number of carriers per beam. Finally, solve the SEC problem by conventional methods (such as Welzl's algorithm or Graham-scan) to find the new beam center locations and beam radius, which are used to calculate the new carrier SNR matrix for the next iteration. We can end the algorithm when the load balancing degree is no longer decreasing.

| **Algorithm 1: Joint Load and Capacity Scheduling** |
|---|
| **Input:** users' coordinates $\{l_n\}$ and beam centers $\{c_k\}$ |
| 1. $i \leftarrow 0$, Initialize the current load balancing degree $\kappa_0$ and carrier SNR |
| 2. **While** $\kappa_{i+1} < \kappa_i$ **do** |
| 3. $\quad$ Solve the problem with *C*1 and *C*2 |
| 4. $\quad$ **Return** $\{R_k^{off}\}$ according to Eq.(4) |
| 5. $\quad$ Adapt $\{x_{k,n}\}$ by sorting $\{R_k^{off}\}$ |
| 6. $\quad$ Update $\{M_k\}$ according to $\{\upsilon_{n,k}\}$ and $\{x_{k,n}\}$ |
| 7. $\quad$ Update $\{c_k\}$ and $\{r_k^i\}$ using Welzl's algorithm |
| 8. $\quad$ Update carrier SNR according to Eq.(1),(2) |
| 9. $\quad$ Update $\kappa$ according to Eq.(10) |
| 10. $\quad$ $i \leftarrow i+1$ |
| 11. **End While** |
| **Output:** $\{R_k^{off}\}$, $\{M_k\}$, $\{c_k\}$, $\{r_k\}$, $\{x_{k,n}\}$ |

## 3.2 Load Scheduling by Flexible Mapping

For comparing with MAP in literature[16] and benchmarking, a strategy with only load scheduling is also assumed. Due to the fixed bandwidth allocation, the constraint *C*1 in should be changed to *C*1' with $W_k \leq M \cdot \tilde{W}, \forall k$. The iterative algorithm is summarized in Algorithm 2. This solution illustrates that the asymmetric scenario without the flexible repeater can benefit from adaptive beam service range based on flexible beam-user mapping.

| **Algorithm 2: Load Scheduling by Flexible Mapping** |
|---|
| **Input:** users' coordinates $\{l_n\}$ and beam centers $\{c_k\}$ |
| 1. $i \leftarrow 0$, Initialize the current load balancing degree $\kappa_0$ and carrier SNR |
| 2. **While** $\kappa_{i+1} < \kappa_i$ **do** |
| 3. $\quad$ Solve the problem with *C*1' and *C*2 |
| 4. $\quad$ **Return** $\{R_k^{off}\}$ according to Eq.(4) |
| 5. $\quad$ Adapt $\{x_{k,n}\}$ by sorting $\{R_k^{off}\}$ |
| 6. $\quad$ Update $\{c_k\}$ and $\{r_k^i\}$ using Welzl's algorithm |
| 7. $\quad$ Update carrier SNR according to Eq.(1),(2) |
| 8. $\quad$ Update $\kappa$ according to Eq.(10) |
| 9. $\quad$ $i \leftarrow i+1$ |
| 10. **End While** |
| **Output:** $\{R_k^{off}\}$, $\{c_k\}$, $\{r_k\}$, $\{x_{k,n}\}$ |

## 4 Performance Evaluation

In the following, we compare the performance of the proposed strategies with exiting works making different efforts in the last five years, which are presented in Table 1. A comparison of different strategies' flexibility is shown in Table 2. These strategies are applied to the model in 4.1 with the parameters in Table 3.

**Table 1 Comparison of Different Strategies' Optimization Efforts**

| Strategy | Label | Optimization Effort |
|---|---|---|
| Flexible Beam Power & Bandwidth[6] | BW-POW | capacity transfer |
| Flexible Beam-User Mapping[16] | MAP | load transfer |
| Flexible Beam-User Mapping with Flexible Bandwidth[16] | BW-MAP | capacity & load transfer |
| Flexible Service Range (proposed) | SR | load transfer |
| Flexible Service Range with Bandwidth (proposed) | BW-SR | capacity & load transfer |

**Table 2 Comparison of Different Strategies' Flexibility**

| Label | Flexibility | | | | |
|---|---|---|---|---|---|
| | carrier power | carrier bandwidth | carriers' number per beam | user mapping | service range & location |
| BW-POW | ✓ | ✓ | ✓ | | |
| MAP | | | | ✓ | |
| BW-MAP | | | ✓ | ✓ | |
| SR | | | | ✓ | ✓ |
| BW-SR | | | ✓ | ✓ | ✓ |

Table 3 System Parameters

| Satellite Parameters | |
|---|---|
| Diagram pattern | Bessel |
| Number of beams | 6 |
| Capacity of considered beams | 6.8Gbps |
| Uniform beam radius | 50km |
| Maximum beam radius $r_f$ | 70.962km |
| Maximum antenna gain | 52 dBi |
| Average free space losses | 210 dB |
| Average atmospheric losses | 0.4 dB |
| Max available bandwidth $W^{total}$ | 500 MHz |
| Polarization | Dual |
| Frequency reuse scheme | 4-Color |
| Number of carriers per color $M$ | 4 |
| Frequency band [GHz] | 20 |
| **User Parameters** | |
| Demand of every active user | 25 Mbps |
| Active users of considered beams | 272 |
| **Receiver Parameters** | |
| Terminal G/T | 16.25dB/K |
| Losses due to terminal depointing | 0.5 dB |

## 4.1 Traffic Model

In order to compare with the strategies proposed in , first, the Dirichlet distribution $Dir(K, \pmb{\alpha})$ with $\pmb{\alpha} = [\alpha_1, \ldots, \alpha_K]$ is adopted to model different traffic profiles while ensuring that the total capacity is $T$; the $\pmb{\alpha}$ feature the possible statistically different traffic demands across beams. 200 Monte-Carlo realizations are run to fulfil random demand variations in two scenarios:

(1) Homogeneous Traffic (HT): Make every possible users' distribution, i.e. different traffic profiles appear with equal probability among all beams, with $\pmb{\alpha} = [1,1,1,1,1,1]$.

(2) Wide Hot-Spot (WHS): Because demand is spatially related, hot spots tend to cluster. Therefore, an imbalanced scenario with two badly overloaded beams surrounded by the colder is explored, with $\pmb{\alpha} = [4,4,1,1,1,1]$. Finally, to obtain more realistic and representative results, we employ a real traffic demand scenario based on population. As shown in Fig. 12(a), unlike the relatively balanced European population distribution in previous studies, the region studied in this paper spans the Heihe-Tengchong line and includes the densely populated provincial capital cities of Chengdu and Chongqing, as well as the sparsely populated plateau region, thus the traffic demand is extremely uneven:

(3) Real Traffic (RT): Apply 64 beams covering the part region of China, 26.64°N~34.99°N, 100.84°E~107.13°E[22]. The coverage area is about 690×800= 55,200km$^2$, and the number of active users is 2897. The total capacity of the beams is about 73Gbps, and the random location of the user is determined according to the traffic demand based on the actual population density. Other system parameters are the same as in the previous two scenarios.

## 4.2 Numerical Results

To better reflect the demand satisfaction and compare with other strategies, we employ the following two normalized parameters, which are common metrics found in the literature:

(1) Normalized quadratic unmet (NQU) capacity:

$$NQU = \frac{\sum_{n=1}^{N}(R^{req}(n) - R^{off}(n))^2}{R_{uni}^{off}} \quad (11)$$

$R_{uni}^{off}$ means all carriers are uniformly sent from the satellite with the same power and bandwidth, and users can appear at all locations within the footprint.

(2) Normalized unmet (NU) capacity:

$$NU = \frac{T - \sum_{n=1}^{N} R^{off}(n)}{T} \quad (12)$$

with $T$ the aggregated demand of all users. To control variate in tackling uneven demand problem, we assume the satellite capacity equals to $T$.

In the rest of 4.2, for each scenario, the averaged NQU, NU, and the offered and minimum rate of Monte-Carlo realizations are summarized in tables, and the detailed views for Cumulative Distribution Function (CDF) of metrics are illustrated in figures.

(1) **Homogeneous traffic scenario**

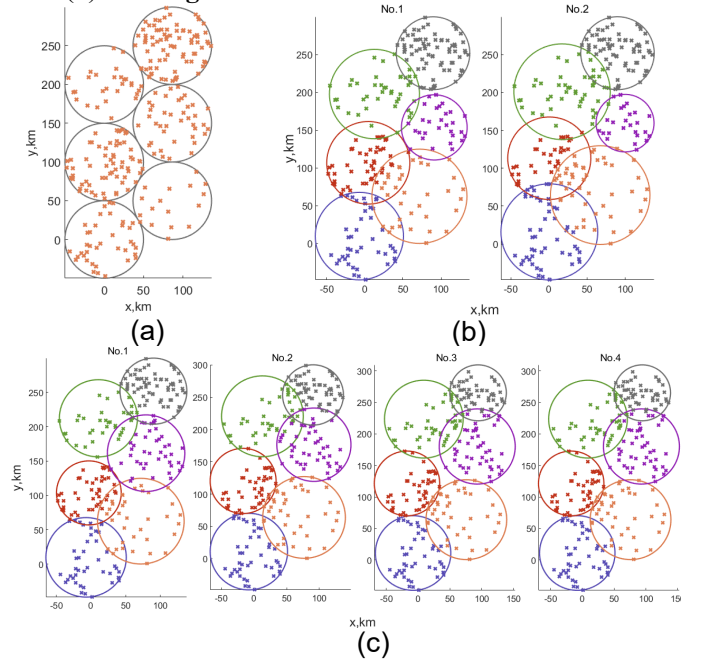

(a)   (b)

(c)

**Fig. 1 Once the whole changing process of flexible beam pattern in HT. (a) Initial uniform pattern. (b) BW-SR: pattern of 3 iterations. (c) SR: pattern of 4 iterations.**

Preliminarily, to fairly evaluate different optimization efforts, all kinds of traffic distributions across beams with equal probability are randomly generated in the HT scenario. Fig. 1 presents once whole changing process of flexible beam pattern and beam-user mapping, where Fig. 1 (a) depicts the initial uniform pattern of 6 beams arranged in two columns and three rows. The circles of different radii represent the service range of the beam, and the "×" marker represents the user in different locations. Except for Fig. 1 (a), the beam is drawn in the same color as the users it serves. Fig. 1 (c) and Fig. 1 (b) depict the changing process of SR and BW-SR, respectively. As each user demands the same rate, it can be seen that after several iterations the loads tend to equalize among all the beams. It is a sub-optimal solution. In addition, with the bandwidth flexibility, for the same users' distribution, the BW-SR needs fewer iteration times than SR.

**Table 4 Strategies' Average Performance in HT Scenario**

| Strategy Label | NQU | NU | Offered Rate [Mbps] | Minimum Rate [Mbps] |
|---|---|---|---|---|
| BW-POW | 0.2039 | 0.3360 | 4518 | 5.750 |
| MAP | 0.0848 | 0.2010 | 5434 | 13.05 |
| BW-MAP | 0.0558 | 0.1645 | 5682 | 12.26 |
| SR | 0.0197 | 0.0924 | 6172 | 20.40 |
| BW-SR | 0.0202 | 0.0870 | 6209 | 16.63 |

In Table 4, it is highly notable that smart mapping provides more improvement in equalizing two-dimensional arrays than one-dimensional lines of beams[16]. This comes from more adjustable space to those congested beams. The flexible beam service range (SR) enhances the performance of user mapping (MAP), even better than bandwidth flexibility (BW-MAP). This is plain to see in Fig. 2, which depicts the NQU in the HT scenario.

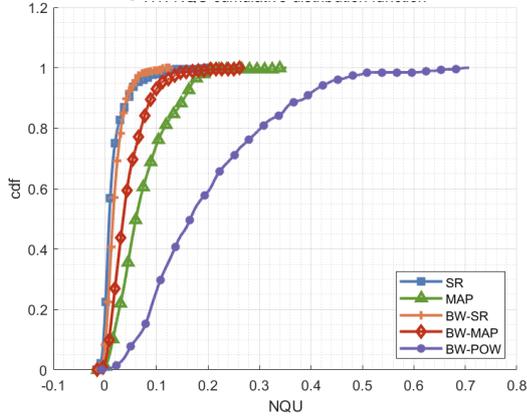

**Fig. 2 CDF of the NQU in HT scenario**

Concerning the offered rates, with similar performance of both SR and BW-SR when pairing together with the flexible mapping and bandwidth, enhancing the performance of the latter by around 8%, as concluded from Table 4. Fig. 3 shows the CDF of the total offered rates for the different strategies relative to the satellite capacity $T$.

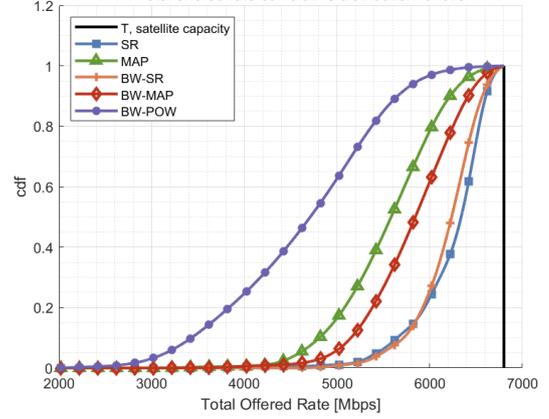

**Fig. 3 CDF of the total offered rate in HT scenario**

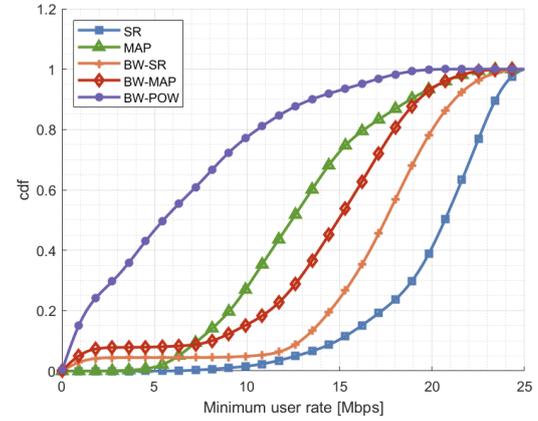

**Fig. 4 CDF of the minimum rates in HT scenario**

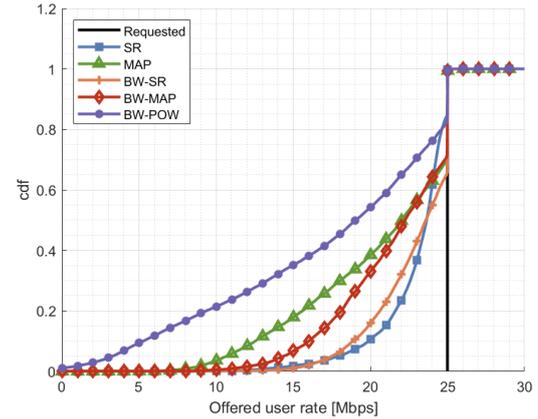

**Fig. 5 CDF of the offered user rates in HT scenario**

The CDF of average minimum rate for all users reflects the fairness of the system, displayed in Fig. 4. On the whole, the flexible service range (SR and BW-SR) achieves lower probabilities for smaller minimum user rates than other strategies. However, due to the strict carrier bandwidth constraints, the optimization performance of only load shifting (MAP and SR) is better than combined capacity and load shifting (BW-MAP and BW-SR, relatively) in the coordinate region where the user rates are close to 0.

Moreover, Fig. 5 shows the CDF of all the offered user rates. Although SR and BW-SR manifest like average total offered rates in Table 4, BW-SR has more users whose rates are in the middle range, both presented in Fig. 3 and Fig. 5.

(2) **Wide hot-spot scenario**

Refer to previous analysis, the joint load and capacity scheduling can address huge strains on the satisfaction of demands in the hot-spot like cases. Therefore, we test the strategies under WH scenario, where two adjacent hot beams surrounded by the colder ones. Table 5 lists the corresponding average performance indices. Particularly, compared with BW-MAP, BW-SR provides a rise of around 26% as for the offered rate, a rise of around 18% as for offered rate, and a reduction of around 77% for NQU, which mean better demand satisfaction and system fairness.

It can be observed that there is no significant change of the average total offered rate of SR and BW-SR, compared with that in HT scenario. In contrast, other strategies have varying degrees of decline in this regard. This reflects the robustness of the two-way effort with the flexible framework.

**Table 5 Strategies' Average Performance in WHS Scenario**

| Strategy Label | NQU | NU | Offered Rate [Mbps] | Minimum Rate [Mbps] |
|---|---|---|---|---|
| BW-POW | 0.1866 | 0.3564 | 4380 | 5.682 |
| MAP | 0.0883 | 0.2191 | 5313 | 12.35 |
| BW-MAP | 0.0776 | 0.2052 | 5047 | 12.97 |
| SR | 0.0145 | 0.0924 | 6173 | 20.84 |
| BW-SR | 0.0175 | 0.085 | 6222 | 17.60 |

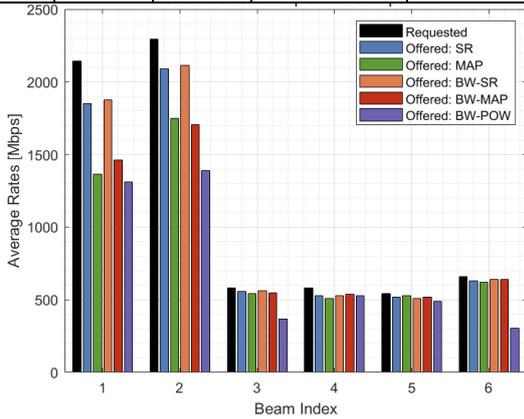

**Fig. 6 Average requested and offered rates per beam**

Fig. 6 shows the average requested and offered rates per beam. Fig. 7 presents the once whole changing process of the flexible beam pattern and beam-user mapping in WHS scenario, where Fig. 7 (a) depicts the initial uniform with two congested ones, index 1 and 2, corresponding to two beams in Fig. 6 with high demand. Fig. 7 (c) and Fig. 7 (b) depict the changing process of SR and BW-SR, respectively.

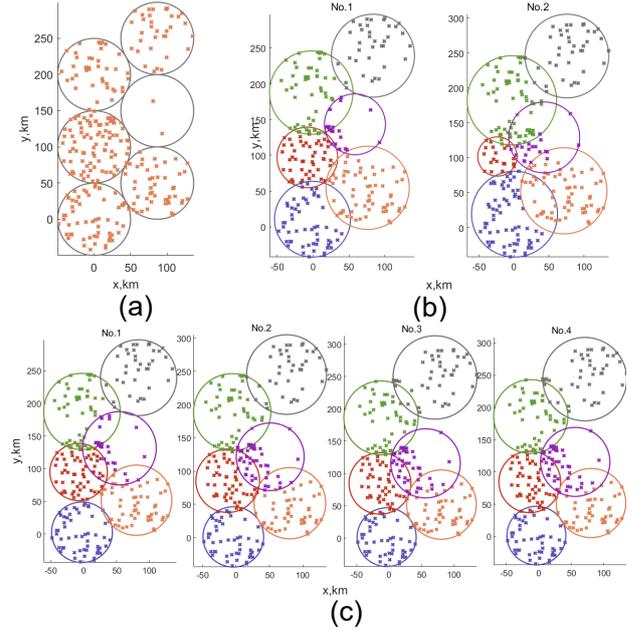

**Fig. 7 Once whole changing process of flexible beam pattern in WHS. (a) Initial uniform pattern. (b) BW-SR: pattern of 3 iterations. (c) SR: pattern of 5 iterations.**

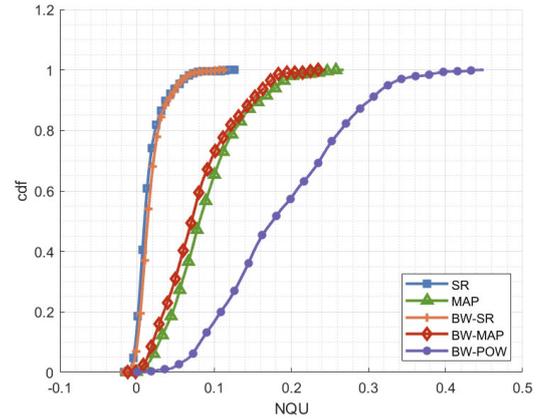

**Fig. 8 CDF of the NQU in WHS scenario**

The CDF of the NQU, total offered rate, minimum rates and offered user rates in WHS scenario are respectively shown in Fig. 8, Fig. 9, Fig. 10 and Fig. 11. Compared with scenario 1, the general trend of distribution probability for these indicators has no major change, but the distinction between different categories of strategies is more obvious. In other words, as for the above 4 metrics, the SR and BW-SR perform similarly and significantly better than MAP and BW-MAP; The latter two are close and much better than BW-POW. Scenario with concentrated hotspots can evaluate the performance of the proposed strategies in dealing with adverse conditions. Note that the SR based on flexible mapping could let all the beams cope with traffic demands through load balancing, while the pure flexible mapping strategy MAP, unable to move beams, failed to achieve complete load balancing. As for rigid mapping strategy BW-POW, due to bandwidth constraints, it was unable to provide sufficient carriers to congested beams, thus further widening the performance gap compared with other strategies.

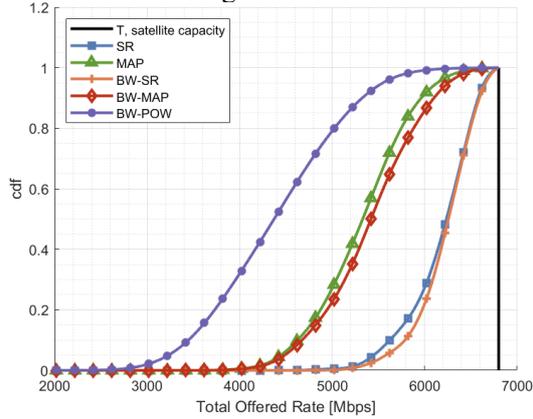

**Fig. 9 CDF of the total offered rate in WHS scenario**

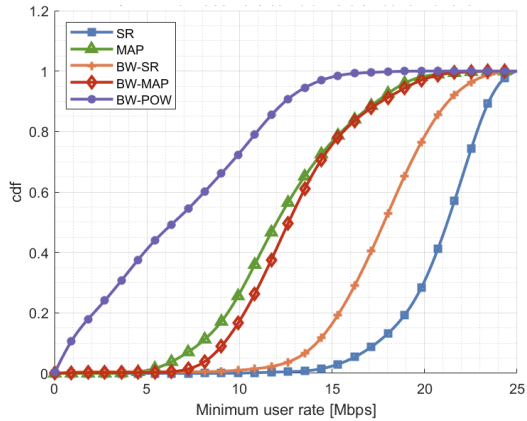

**Fig. 10 CDF of the minimum rates in WHS scenario**

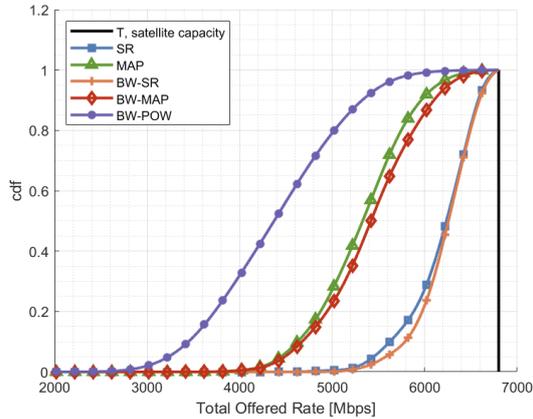

**Fig. 11 CDF of the offered user rates in WHS scenario**

**(3) Real traffic scenario**

Finally, we explore a practical example with 64 beams serving a traffic demand distribution based on part of China's population density as presented in Fig. 12(a). This is significant, as this demand distribution is markedly different from the more evenly distributed European population patterns previously studied. Fig. 12 (b) depicts the initial uniform pattern of 64 beams arranged in 8 columns and 8 rows. Fig. 12 (c) and Fig. 12(d) depict the final results of flexible pattern in RT scenario for the SR and BW-SR, respectively.

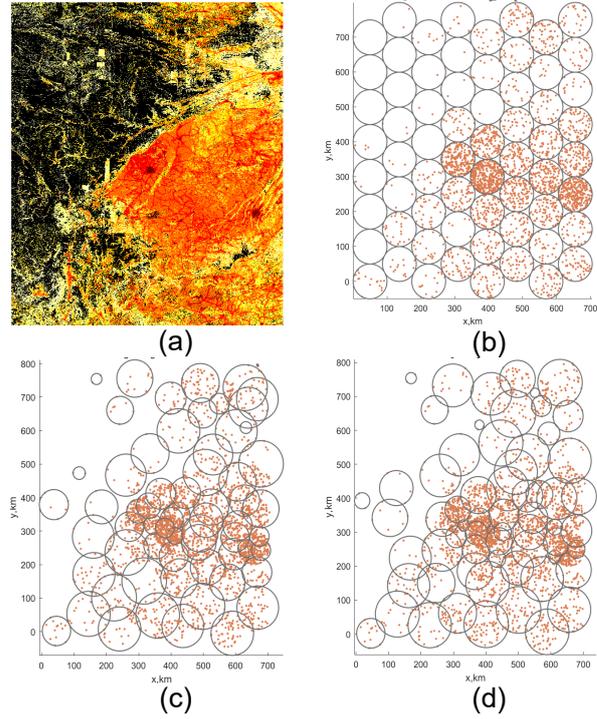

**Fig. 12 Flexible beam pattern results in RT. (a) Corresponding population density map. (b) Initial uniform pattern. (c) SR: pattern of 11th iteration. (d) BW-SR: pattern of 5th iteration**

Table 6 lists the corresponding performance indices of one run in RT scenario. The results demonstrate the inherent robustness of SR and BW-SR, based on load balancing achieved through reconfigurable active antenna, thereby beforehand alleviates the challenge faced by the flexible repeater in severe scenarios.

**Table 6 Strategies' Once Performance in Real Scenario**

| Strategy Label | NQU | NU | Offered Rate [Mbps] | Minimum Rate [Mbps] |
|---|---|---|---|---|
| BW-POW | 0.8186 | 0.8806 | 8641 | 0 |
| MAP | 0.2714 | 0.4321 | 41129 | 2.531 |
| BW-MAP | 0.2744 | 0.4332 | 41049 | 2.498 |
| SR | 0.0757 | 0.2524 | 54141 | 15.46 |
| BW-SR | 0.0620 | 0.2063 | 57481 | 11.50 |

Fig. 13 shows the average requested and offered rates per beam in RT scenario. It can be seen that the traffic demands of different beams vary significantly in order of magnitude, with a few spare beams. In this case, it is hard for BW-POW to meet the traffic needs of the congested beams. Fig. 14 presents the CDF of the minimum rate for all users. Consistent with previous experiments, the improvement provided by strategies based on flexible service range is significant. In the above three scenarios, it appears that apart from a reduced number of iterations, SR-BW does not exhibit a discernible advantage over SR. However, we should keep in mind that in order to control variables, the current research is limited to addressing the uneven spatial distribution of demand. As traffic demand evolves over time, particularly in situations where total demand exceeds total supply, the multi-dimensional resource scheduling strategy is expected to confer greater advantages.

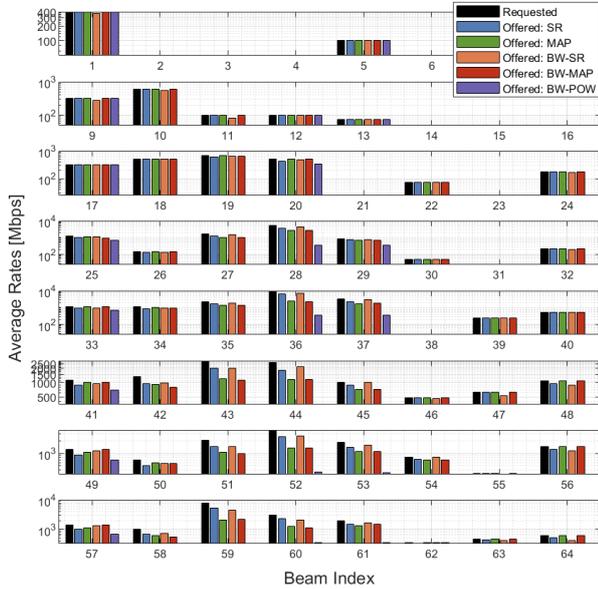

**Fig. 13 Average requested and offered rates per beam**

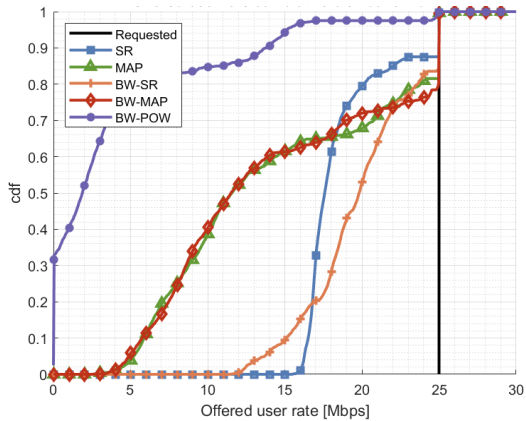

**Fig. 14 CDF of the offered user rates in RT scenario**

The MATLAB implementation time spent on a single run in the case of the example presented above, and the average consuming time in HT and WHS scenario of SR and BW-SR are collected in Table 7. The running time of the proposed strategy is similar in HT and WHS scenarios. Due to the bandwidth flexibility, adopting BW-SR takes about 70% of the time relative to SR, with a little lower performance caused by carrier constraints $C1$ and $C2$, as discussed before. All the simulations run on a commercial laptop with Gen Intel(R) Core (TM) i9-13900HX and 32 Gigabytes of RAM. It proves that, with the increase of both user and beam, the proposed strategies have fine convergence in general. In practice, the strategy does not always run from an initial uniform state, just when the traffic demand changes markedly, thus it leads to a good tradeoff between demand satisfaction and implementation.

**Table 7 Time-Consuming in 3 Scenarios**

| Scenario | HT | | WHS | | RT | |
|---|---|---|---|---|---|---|
| Beam Number | 2×3=6 | | 2×3=6 | | 8×8=64 | |
| User Number | 272 | | 272 | | 2987 | |
| Strategy | SR | BW-SR | SR | BW-SR | SR | BW-SR |
| Iteration Number | 4.98 | 3.47 | 5.40 | 3.64 | 11 | 5 |
| Running Time [min] | 1.11 | 0.71 | 0.91 | 0.70 | 76.3 | 37.8 |

As a closing thought, we should note that the performance evaluation in this work relies on one GEO satellite, leaving it to future research to extend this strategy to NGSO constellations. Additionally, full-frequency reuse can be employed to significantly enhance the spectral efficiency of the system, by using advanced precoding techniques[4, 23] to mitigate inter-beam interference. Finally, we believe that a DBF (digital beamforming)-based multi-dimensional resource management as per the traffic requirement, by further optimizing the combination of the transmission power and beam gain for maximizing the traffic accommodation rate.

## 5  Conclusion

This work has explored an optimization strategy embodying two-way effort for the reconfigurable active satellite antenna, which combines capacity transfer and load transfer to serve non-uniform traffic demands in HTS footprints. Guided by flexible beam-user mapping, a simple iteration for the complex beam optimization problem has revealed as a practical strategy. Both Dirichlet distribution and a real traffic demand based on population has been chosen to test performance. From the results, it can be concluded that the proposed strategies operate better than joint power-bandwidth allocation, and results in a good tradeoff between system unmet capacity and implementation complexity. Notably, for those cases for which traffic is extremely imbalanced across beams, the use of a two-way effort can report significant benefits when serving the requested traffic levels.

### Acknowledgment

The authors would like to express their gratitude to Dr. Tomás Ramírez, Dr. Carlos Mosquera, and Dr. Nader Alagha for their valuable support. This work was supported in part by the Aerospace Discipline Education New Engineering Project, grant number 145AXL250004000X.

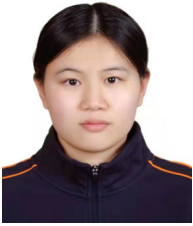

**ZHUOYA JIA** received a B.S. degree from Space Engineering University, Beijing, China, in 2022, where she is currently pursuing the master's degree in the National Key Laboratory of Space Target Awareness. Her research interests include the analysis and integration of space-based information systems, satellite resource scheduling and array signal processing.

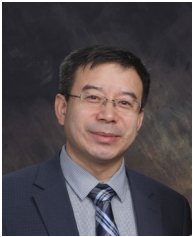

**Wei Xiong** was born in 1971. He received his Ph.D. degree at Beihang University, in 2005. He is currently a professor in the National Key Laboratory of Space Target Awareness, Space Engineering University. His research interests include complex networks, systems engineering and theory.

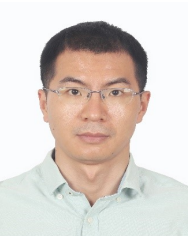

**Hongxing Hao** was born in 1987, and is now a research assistant in Space Engineering University, Beijing, China. He received his doctor's degree in systems engineering in 2014. His research interests include sparse representation of complex values, estimation of interferometric images, remote sensing image processing, system science and so on.

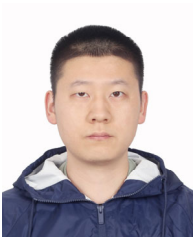

**Zhen Liu** was born in 1995. He received B.S. degree from the Space Engineering University in 2018, where he is currently pursuing the Ph.D. degree. His research interests include system modeling, system evaluation and deep reinforcement learning.

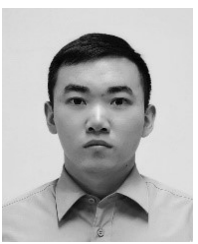

**Chi Han** was born in 1997. He received the M.S. degree from the Science and Technology on the National Key Laboratory of Space Target Awareness, Space Engineering University, in 2021, where he is currently pursuing the Ph.D. degree. His current research interests include satellite communication and spatial information system analysis.